%% file: ms.tex
\documentclass[fleqn,10pt]{wlscirep}
\pdfoutput=1
\usepackage[utf8]{inputenc}
\usepackage[T1]{fontenc}
\usepackage{textcomp}
\usepackage{gensymb}

\graphicspath{ {./figures/} }

\title{Persistent meanders and eddies lead to a quasi-steady Lagrangian transport pattern in a weak western boundary current}

\author[1,*]{M. B. Gouveia}
\author[2,3]{R. Duran}
\author[1]{J. A. Lorenzzetti}
\author[4]{A. T. Assireu}
\author[5]{R. Toste}
\author[5]{L. P. de F. Assad}
\author[1]{D. F. M. Gherardi}

\affil[1]{Division of Remote Sensing, Brazilian National Institute for Space Research, São José dos Campos, 12227-010, Brazil}
\affil[2]{National Energy Technology Laboratory, Albany, OR 97321, USA}
\affil[3]{Theiss Research, La Jolla, CA, 92037, USA}
\affil[4]{Institute of Natural Resources, Federal University of Itajubá, Itajubá, 37500-015, Brazil}
\affil[5]{Laboratory for Computational Methods in Engineering, COPPE/UFRJ, Rio de Janeiro, 21941-907, Brazil}

\affil[*]{mainarabg@gmail.com}

\include{sections/00_abstract}

\begin{document}

\flushbottom
\maketitle
%
%
\thispagestyle{empty}

\section*{Introduction}
\input{sections/00_intro}

\section*{Results}
\input{sections/01_results}

\section*{Discussion}
\input{sections/02_discussion}

\section*{Summary and Conclusions}
\input{sections/03_conclusions}

\section*{Data and Methods}
\input{sections/04_methods}

\bibliography{bib/references.bib}

\section*{Acknowledgements}
\input{sections/10_acknowledgements}

\section*{Author contributions statement}
\input{sections/11_author-contributions}

\section*{Competing interests}
\input{sections/12_competing-interests}


\end{document}

%% file: sections/00_abstract.tex
\begin{abstract}

The Brazil Current (BC) is a weak western boundary current flowing  along the Southwestern Atlantic Ocean.
It is frequently described as a flow with intense mesoscale activity and relatively low volume transport between 5.0 to 10.0 Sv. 
We use a 13-year eddy-resolving primitive-equation simulation to show that the presence of persistent meanders and eddies leads to characteristic quasi-steady Lagrangian transport patterns, aptly extracted through climatological Lagrangian Coherent Structures (cLCSs).
The cLCSs position the surface expression of the BC core along the 2000 $m$ isobath,  in excellent agreement with high resolution satellite sea-surface temperature and the model Eulerian mean velocity.
The cLCSs deformation pattern also responds to zonally persistent cross-shelf SSH transition from positive (high) values near coastline to low (negative) values between 200 and 2000 $m$ and back to positive (high) offshore from the 2000 $m$ isobath. 
Zonally-paired cyclonic and anticyclonic structures are embedded in this transition, also causing the cLCSs to deform into chevrons. 
An efficient transport barrier is identified close to the 200 $m$ isobath confirmed by limited inshore movement of drogued buoys and accurately indicated by an along slope maxima of climatological strength of attraction.
We also show that the persistent cyclonic and anticyclonic structures may induce localized cross-shelf transport. 
Regions of low climatological strength of attraction coincide with large shelves and with stagnant synthetic trajectories. 
We also show that cLCSs accurately depict trajectories initiated at the location of Chevron’s spill (November 2011) as compared to synthetic and satellite trajectories, and the outline of the oil from that accident. 
There is also an agreement between the large-scale oil slicks reaching the Brazilian beaches (from August 2019 to February 2020) and the strength of climatological attraction at the coast.
The identification and quantitative description of climatological Lagrangian coherent structures is expected to improve the effectiveness of future emergency response to oil spills, contingency planning, rescue operations, larval and fish connectivity assessment, drifter launch strategies, waste pollutant dispersion and destination. Our work also clarifies the influence of persistent mesoscale structures on the regional circulation.

\end{abstract}


%% file: sections/00_intro.tex
Subtropical western boundary currents (WBC) are one of the main contributors to the meridional ocean transport of heat and salt \cite{wiggins2005dynamical,assad2009volume,imawaki2013western}.
They are generally depicted as continuous surface flows and their intensity and persistence are considered as a dynamic barrier for cross-flow transport \cite{stommel1957survey}, influencing the pathway of pollutants and fish larvae \cite{deOliveira2018three,daRochaFragoso20164d,dagostini2015connectivity,marta2013efficient,romero2013integrated}.
However, the Brazil Current (BC) is considered a weak WBC in the Southwestern Atlantic Ocean (SWA) \cite{peterson1991upper}.
The BC transports a relatively small volume (values ranging 4--6.5 Sv) \cite{castro2006estrutura} of water southward in the upper 520 $m$ \cite{stramma1999water}.
Strong mesoscale activity develops in the vicinity of the BC, mainly near 22S, with large frontal meanders and eddies \cite{silveira2008meander}.
Despite being weak relative to other western boundary currents, the BC has been found to exert significant control over Lagrangian transport \cite{gouveia2017brazilian}, in this paper we examine persistent Lagrangian transport patterns in a WBC with persistent meanders and eddies. 

The baroclinically unstable nature of the BC is caused by the presence of the Intermediate Western Boundary Current flowing to the north between 800 and 1000 $m$ depth, below the BC, which in the upper 500 $m$ transports the Tropical Water (TW) and the South Atlantic Central Water to the south \cite{silveira2008meander}.
The formation of meanders and eddies can also locally reverse the current flow offshore of the 1000 $m$ isobath and cause changes in current transport \cite{lima2016assessment}.
The intense mesoscale activity along the BC, from its origin in the bifurcation of the South Equatorial Current down to the Brazil-Malvinas confluence region, includes recurrent or semi-permanent meanders, cyclonic and anticyclonic structures, and eddies \cite{campos1995water,campos1996experiment}. 

The method to compute climatological Lagrangian Coherent Structures (cLCSs) was developed recently and has been used to extract important Lagrangian transport patterns from large velocity time series \cite{duran2018extracting}.
Pattern identification include 1) isolated regions where trajectories are unlikely to leave or enter; 2) regions that attract nearby parcels of water and therefore are more susceptible to pollution impacts; and 3) recurrent transport patterns.
Some recent studies have shown the relevance of cLCSs.
\cite{gough2019persistent} showed the efficacy of some of the cLCSs as transport barriers by using synthetic drifters advected by the instantaneous model velocities and by using 3207 satellite-tracked drifter trajectories spanning over two decades (1994--2016).
\cite{maslo2020connectivity} showed that cLCSs were efficient in identifying predominant transport patterns in the deep ($\approx$ 1500 $m$) Gulf of Mexico, as determined by RAFOS floats and synthetic drifter trajectories. 

To compute cLCSs, \cite{gough2019persistent} and \cite{maslo2020connectivity} used a free-run simulation performed with NEMO (Nucleus for European Modelling of the Ocean) and ROMS (Regional Ocean Modeling System), respectively, while \cite{duran2018extracting} used an operational HYCOM (Hybrid Coordinate Ocean Model) simulation. 
Thus, in combination, these three papers show that cLCSs are robust in bypassing the variability inherent to geophysical flows, while accurately identifying predominant transport patterns. Searching for structures that were able to bypass the chaotic nature of transport while extracting predominant, and important, transport patterns from long velocity time series was the motivation behind the development of cLCSs \cite{duran2018extracting}.

In this paper we show that cLCSs are also efficient in extracting the predominant circulation from another free-running numerical simulation, in the distinct setting of a WBC, characterized by the presence of persistent and recurrent eddies and meandering.
In this paper we offer new insights on the interpretation of cLCSs by relating them to the time-mean structure of Eulerian fields such as  satellite sea-surface temperature (SST), model sea-surface height (SSH), and eddy-, mean-, and total-kinetic energies (EKE, MKE, and TKE, respectively).
Further interpretation of cLCSs is based on comparisons with satellite-tracked drifters and synthetic drifter trajectories. 
The transport patterns associated with persistent Eulerian structures include regions with increased Lagrangian variability and offshore transport.
We identify a cross-shelf transport barrier, separating distinct dynamical regimes, and persistent recirculation patterns in the transition region between them. 
We also show that cLCSs highlight the transport patterns that help explain the observed drift of the Chevron's oil spill (November of 2011) and the recent large-scale oil slicks observed at Brazilian beaches (from August 2019 to February 2020).
Here, cLCSs are computed from daily-mean surface velocities from a 13 year (2003--2015) ROMS simulation \cite{shchepetkin2005regional,shchepetkin2009correction} with an eddy-resolving grid of $1/36\degree$ ($\approx$ 3 $km$) and 40 vertical levels.
Following the method described in \cite{duran2018extracting}, we compute 7-day sliding window Cauchy-Green Tensors (CGT) in reverse time, from the daily climatological velocity.
From these CGT we calculate the monthly and yearly average CGT.

%% file: sections/01_results.tex
The underlying causes of quasi-steady attracting Lagrangian structures may differ.
It is therefore important that the physical interpretation of monthly climatological attraction strength ($c\rho$) and cLCSs are supported by the combined use of complementary data.
In this study we use satellite data, \textit{in situ} observations and numerical model outputs.

\subsection*{The mean structure of the BC}

\begin{figure}
    \centering
    \includegraphics[width=0.9\textwidth]{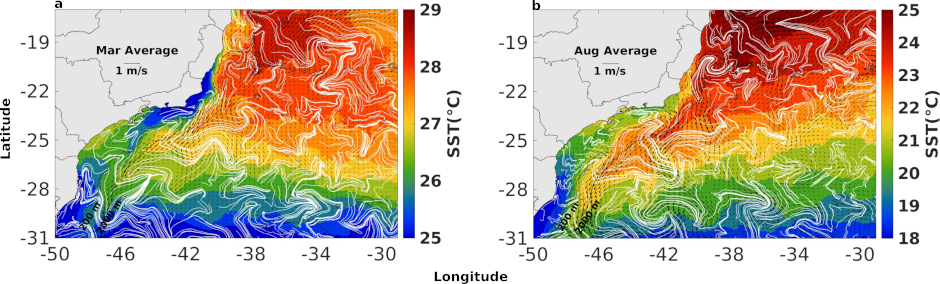}
    \caption{March (a) and August (b) monthly mean SST from the MUR dataset (color shading), with spatial resolution of $0.01\degree$ provided by PODAAC (Physical Oceanography Distributed Active Archive Center), superposed by the corresponding monthly-mean-surface ocean velocity from ROMS (black arrows) and cLCSs (white lines). Depth contours of 200 and 2000 $m$ are indicated in black lines. Note that color bars are different for each month.}
    \label{fig:01:mur}
\end{figure}

The typical mean structure of the BC between 22 and 31S is easily extracted by time averaging Eulerian fields.
The monthly-mean model Eulerian surface flow aligns well with the advection patterns of monthly-mean satellite SST from Multi-scale Ultra-high Resolution (MUR) dataset, suggesting a continuous surface poleward flow along the 2000 $m$ isobath (Fig.~\ref{fig:01:mur}). 

In the austral summer (Fig.~\ref{fig:01:mur}a), between 17--20S and 38--29W cLCSs suggest Lagrangian transport from east to west, in agreement with the Eulerian mean.
Between 20--22S meandering is more common by Vitória-Trindade seamount chain (20S and 39--34W), including over the shelf near 20S where the mean Eulerian flow and cLCSs clearly show a meander (March) or an eddy (August).
South of about 22S, the BC main axis of the flow is usually oriented poleward along the 2000 $m$ isobath, with climatological squeezelines deforming as chevrons, similar to SST chevrons.

Similar chevrons by climatological squeezelines can also be found in a coastal upwelling region, indicated by low SST near the shore in the austral summer (Fig.~\ref{fig:01:mur}a), when there is an intensification of upwelling, although coastal circulation inshore of the 200 $m$ isobath is equatorward.
In the austral winter the chevron shape is observed only near the upwelling Cabo Frio (23S and 42W; Fig.~\ref{fig:01:mur}b). 

The advection of TW by the BC causes a zonal front with the cold and low salinity coastal water, between 21S and 31S in both seasons.
We observed the presence of stronger thermal fronts in the austral summer, due the intensification of upwelling at Cabo Frio (23S) and Cabo de Santa Marta (28S).
In March the TW, between 21 and 30S, is characterized by surface water with temperatures above $26.5\degree C$ (Fig.~\ref{fig:01:mur}a) and in August by temperatures above $21\degree C$ (Fig.~\ref{fig:01:mur}b).

Persistent offshore advection can be seen through satellite SST, cLCSs and monthly Eulerian velocity in austral summer and winter (Fig.~\ref{fig:01:mur}a and \ref{fig:01:mur}b).
In the winter, the Eulerian mean surface velocity shows persistent offshore flow at 19.5S and 39W, around 23S and 39W, between 25--26.5S and 42W (Fig.\ref{fig:01:mur}a). 
In the summer offshore transport is similarly located but can also be seen around 30S and 47W (Fig.~\ref{fig:01:mur}b).
Often these Eulerian offshore patterns coincide with cLCSs deforming as chevrons, and in the summer they coincide with satellite SST advection. 
In the austral summer, SST is consistent with offshore advection (e.g. 26S and 42W), and chevron-like cLCSs (Fig.~\ref{fig:01:mur}a), while in the austral winter offshore advection (e.g. 26S and 42W) does not coincide with offshore SSH advection, yet cLCSs do conform to the offshore Eulerian flow (Fig~\ref{fig:01:mur}b).

\begin{figure} 
    \centering
    \includegraphics[width=0.9\textwidth]{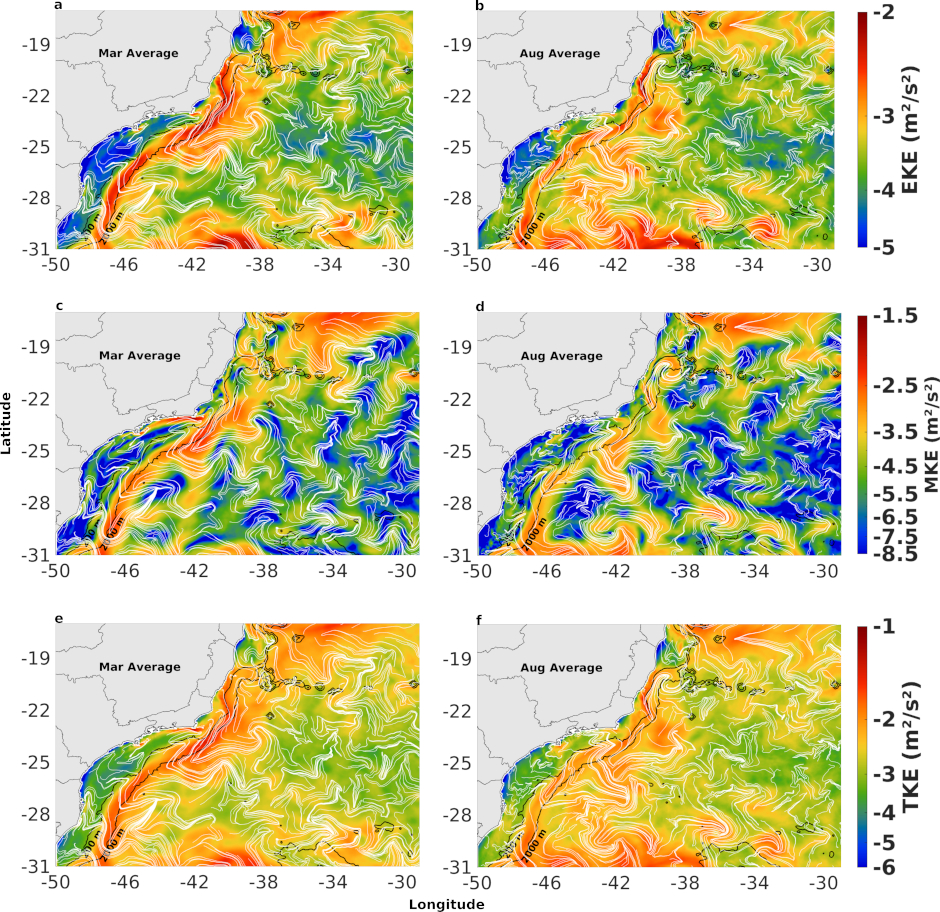}
    \caption{Monthly average maps of EKE (a and b), MKE (c and d) and TKE (e and f) (color shading) for March (left frames) and August (right frames), from 2003 to 2015. Units are $m^2/s^2$ in log scale. Monthly cLCSs are represented by white lines. The 200 and 2000 $m$ depth contours are represented by black lines.}
    \label{fig:02:eke}
\end{figure}

Variability in the BC is evident in austral summer and winter through an intensification of eddy kinetic energy (EKE), between the 200 $m$ and 2000 $m$ isobaths south of about 23S (Fig.~\ref{fig:02:eke}a and~\ref{fig:02:eke}b).
This alongshore EKE maxima is collocated to the Eulerian mean peak BC velocity which is closely aligned with the 2000 $m$ isobath, the mean peak velocity being  in agreement with chevron-shaped cLCSs (Fig.~\ref{fig:02:eke}c and~\ref{fig:02:eke}d).
The alongshore EKE maxima south of about 22S, between the 200 and 2000 $m$ isobaths, is about 0.1 $m^2/s^2$ and slightly more energetic in the Summer (March).
The alongshore MKE maxima is centered along the 2000 $m$ isobath south of about 22S is about 0.1 $m^2/s^2$ in the summer and about 0.05 $m^2/s^2$ in the winter.

In the summer, EKE maxima offshore of the 2000 $m$ isobath, are located between 17 and 20S, offshore of the Abrolhos Bank and between 21 and 26S, the latter is  adjacent to the alongshore EKE maxima, between the 200 and 2000 $m$ isobaths (Fig.~\ref{fig:02:eke}a). 
In the winter, EKE maxima just offshore from the 2000 $m$ isobath can be located between 17--18S,  and near 22.5S, 25.5S and 30S (Fig.~\ref{fig:02:eke}b).

In winter and summer, maxima MKE structures that are just offshore to the mean BC can also be seen, but are more localized near 22.5 and near 25S (Fig.~\ref{fig:02:eke}c and~\ref{fig:02:eke}d).
These locations coincide with offshore flow in Eulerian-mean velocity (Fig.~\ref{fig:01:mur}a and \ref{fig:01:mur}b). 

The low SST associated with intermittent coastal upwelling off Cabo Frio during austral summer (23S and 41--46W; Fig.~\ref{fig:01:mur}a) coincides with a coastal jet in the Eulerian-mean velocity which deforms cLCSs as chevrons, and can be seen as an MKE maxima (Fig.~\ref{fig:02:eke}c).

\begin{figure} 
    \centering
    \includegraphics[width=0.9\textwidth]{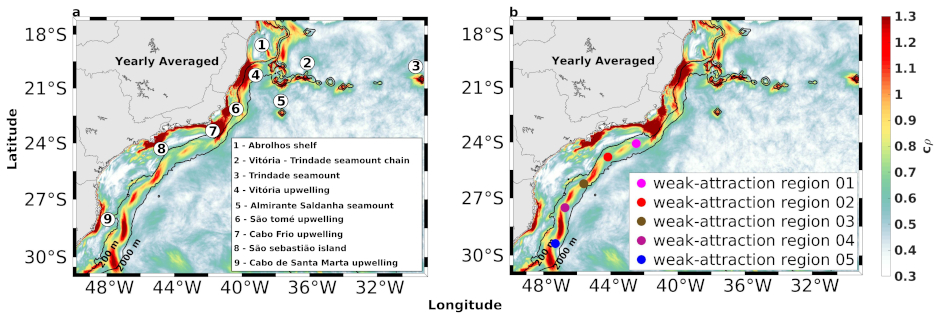}
    \caption{Climatological attraction $c\rho$ (colors, logarithmic scale), computed by yearly averaging 7-day CG tensors from the 2003–2015 climatology,  showing (a) locations with high values of annual mean $c\rho$ ($> 1, 2.7$ in linear units) and (b)  regions of weak attraction adjacent to the 2000 $m$ isobath (colored dots).The 200 and 2000 $m$ depth contours are represented by black lines.}
    \label{fig:03:attraction}
\end{figure}

High stretching values ($c\rho$ $> 1$  in logarithmic scale, or $> 2.7$ in linear units) are found near steep bathymetry areas such as seamounts (Regions 1, 2, 3 and 5 in Fig.~\ref{fig:03:attraction}a) and upwelling regions (Regions 4, 6, 7, and 9 in Fig.~\ref{fig:03:attraction}a). 
Medium to high stretching values ($c\rho$ > 0.7, or > 2 in linear units) are found adjacent to the BC core between 200 to 2000 $m$ isobaths, and between 23--31S with regions of strong attraction interspersed by weakly-attracting regions (Fig.~\ref{fig:03:attraction}). 
Along the slope, climatological attraction strength extrema coincides with EKE extrema, including a minimum near region 1 (Fig.~\ref{fig:02:eke} and~\ref{fig:03:attraction}).

\begin{figure} 
    \centering
    \includegraphics[width=0.9\textwidth]{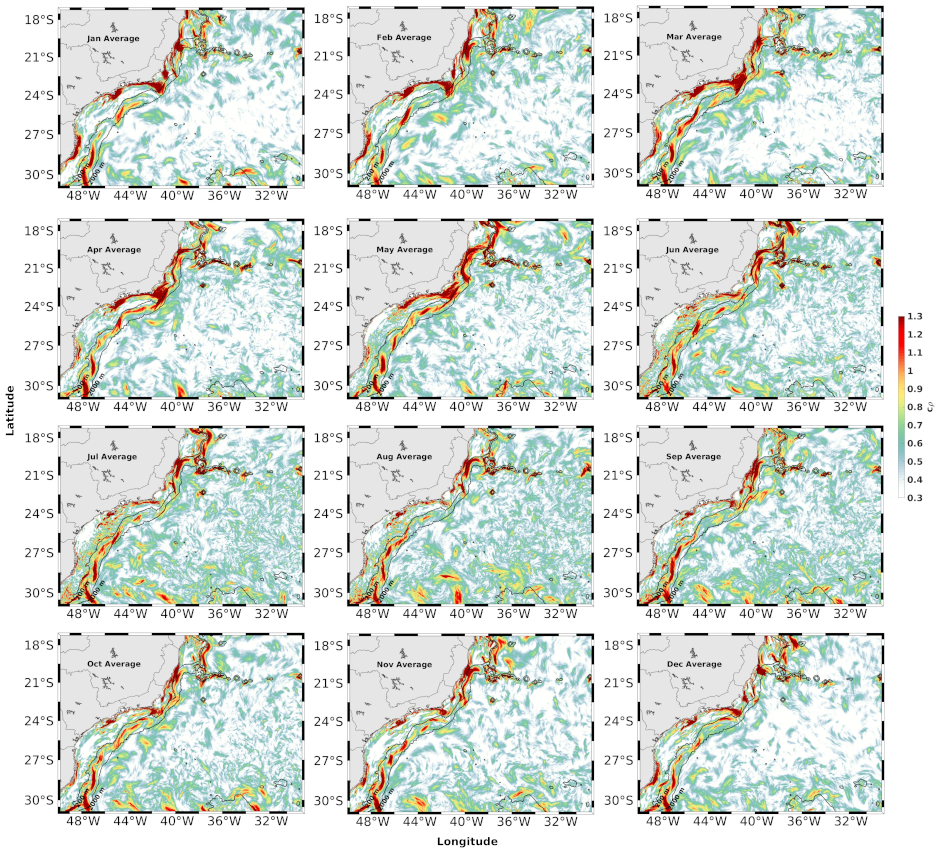}
    \caption{Monthly climatological attraction strength ($c\rho$) in the SWA. Note the discontinuous meridional distribution of high (red) $c\rho$ values between the 200 and 2000 $m$ depth contour lines (thin black lines).}
    \label{fig:04:crho}
\end{figure}

The climatological attraction strength, $c\rho$ shown in Fig.~\ref{fig:04:crho}, has some variability over the slope (200 to 2000 $m$ isobaths) between 23S and 31S. 
This variation in $c\rho$ suggests low-frequency variability of Lagrangian transport patterns with a tendency for some cross-slope transport where low $c\rho$ predominates (see Supplementary Fig. S1 and S2). 
In general, the low frequency time variability suggested by $c\rho$ over the slope, near the latitudes 23.5--24.5S and between 26--27S, is largest from December to February.
In March, June, July, August and September high values of $c\rho$ can be seen along most of the slope south of 23--25S, in contrast with other months where minima are interspersed with maxima. 
Continuous high values of $c\rho$ over the slope are observed in April, May, October and November months. 
The coastline and shallow shelf between 24 and 27S has a $c\rho$ minima in all months, sometimes contrasting with a strong maxima along the coastline north of 24S (e.g. January through May). 
The $c\rho$ minima in the nearshore environment and coastline between 24 and 27S identifies a stagnant region through the year, a region that should be relatively safe from spills originating outside the $c\rho$ minima (Supplementary Fig. S3, middle panel). 
However, any pollution originating within this region including the coastline is unlikely to disperse, thus possibly causing a greater impact due to a higher concentration of contaminants (Supplementary Fig.S3, right panel).

An intensification of $c\rho$ near Cabo Frio upwelling region (Region 7 in Fig.~\ref{fig:03:attraction}a) is observed from January to May due to an increase in speed associated with a coastal jet (Fig. S4) between São Sebastião Island to Cabo Frio (Region 8 to 7 in Fig.~\ref{fig:03:attraction}a).
A peculiar $c\rho$ maxima with "U" shape can be seen in the months of April and May when the surface speed bends cyclonically between 23 to 24S and 44 and 41.5W (Fig.~\ref{fig:04:crho}, see also Fig. S4). 

\begin{figure} 
    \centering
    \includegraphics[width=0.9\textwidth]{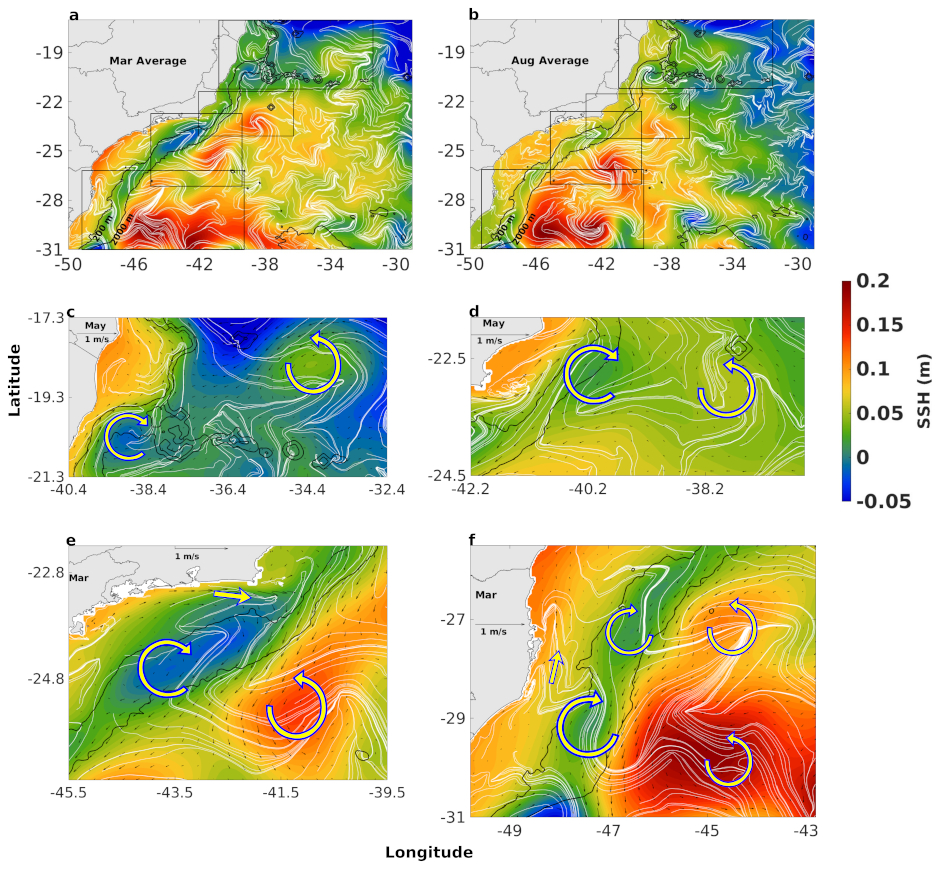}
    \caption{March (a) and August (b) monthly climatological squeezelines (white lines) over model monthly-mean SSH. Black rectangles indicate regions shown in panels (c) to (f). (c) Abrolhos Bank region (location 1 in Fig.~\ref{fig:03:attraction}a), (d) Cabo de São Tomé (location 6 in Fig.~\ref{fig:03:attraction}a), (e) Cabo Frio (location 7 in Fig.~\ref{fig:03:attraction}a) and (f) Cabo de Santa Marta (Region 9 in Fig.~\ref{fig:03:attraction}a). The presence of cyclonic and anticyclonic eddies features are highlighted by curved arrows. Straight arrows indicate the direction of coastal jets at upwelling regions. Reference months are indicated at the top left of each map. The vector scale is on the top of each map. Depth contours of 200 and 2000 $m$ are represented by thin  black lines.}
    \label{fig:05:ssh}
\end{figure}

A description of the mean structure BC is organized in four different regions based on structures from the model-mean SSH to illustrate upwelling process and persistent mesoscale activity and their association with persistent Lagrangian transport patterns as seen through cLCSs  (Fig.~\ref{fig:05:ssh}).

The first region (Fig.~\ref{fig:05:ssh}c) comprises a group of zonally oriented topographic features near 20S starting at the Abrolhos shelf on the west (location 1 in Fig.~\ref{fig:03:attraction}a), the Vitória-Trindade Seamount chain (location 2 in Fig.~\ref{fig:03:attraction}a), and the oceanic island complex of Trindade Martin Vaz (location 3 in Fig.~\ref{fig:03:attraction}a) on the east. 
An important persistent oceanic feature captured by cLCSs in this region is the Vitória cyclonic eddy to the south of Abrolhos shelf in addition to the Abrolhos anticyclonic eddy on the east centered around 34.4W (Fig.~\ref{fig:05:ssh}c). 
The Abrolhos anticyclonic eddy translates from east to west near Vitória-Trindade seamount chain (Region 2 in Fig.~\ref{fig:03:attraction}a) between the latitudes 18 and 20S and longitudes of 38 and 32W (see Fig. S5). 
The second region (Fig.~\ref{fig:05:ssh}d) is characterized by the presence of a coastal upwelling near the Cabo de São Tomé (location 6 in Fig.~\ref{fig:03:attraction}a), identified by low SST values in austral summer (Fig.~\ref{fig:01:mur}a), just offshore from the upwelling there is a cyclonic structure centered around 22S and 40.2W.
Further offshore there is an anticyclonic structure centered at 23S and 38W. 
Both of these structures induce persistent Lagrangian transport seen through the deformation of cLCSs. 
This region also coincides with EKE and MKE maxima (Fig.~\ref{fig:02:eke})  protruding offshore from the 2000 $m$ isobath, and with an Eulerian mean velocity (Fig.~\ref{fig:05:ssh}d) all of which suggest offshore transport.
A synthetic drifter experiment confirms that the persistent mesoscale structures are likely to cause considerable cross-shelf transport (see Supplementary Fig. S2a--b).
A third region (Fig.~\ref{fig:05:ssh}e) is located in front of the east-west oriented coast of Rio de Janeiro (location 7 in Fig.~\ref{fig:03:attraction}a) with a coastal upwelling jet near Cabo Frio deforming cLCSs.
A couple of cyclonic and anticyclonic features between 23--25S, and the 200 and 2000 $m$ isobaths cause an onshore-offshore-onshore-offshore sequence, although cross-shelf transport seems limited not passing the 2000 $m$ isobath where the core of the BC can be seen except possibly near 23S.
Further offshore near 41W and 25S, there is a SSH maxima associated with recurrent anticyclonic Lagrangian flow depicted by cLCSs deformation (Fig.~\ref{fig:05:ssh}e, see also Fig. S5). 
Near Cabo de Santa Marta (location 9 in Fig.~\ref{fig:03:attraction}a) is the fourth and southernmost region (Fig.~\ref{fig:05:ssh}f), where the surface flow is influenced by a distinctive coastal upwelling region in 29S, and the presence of two dipole-like structures, suggesting cross-shelf variations over the 200 $m$ isobath, restricted inshore of the 2000 $m$ isobath.

\begin{figure} 
    \centering
    \includegraphics[width=0.9\textwidth]{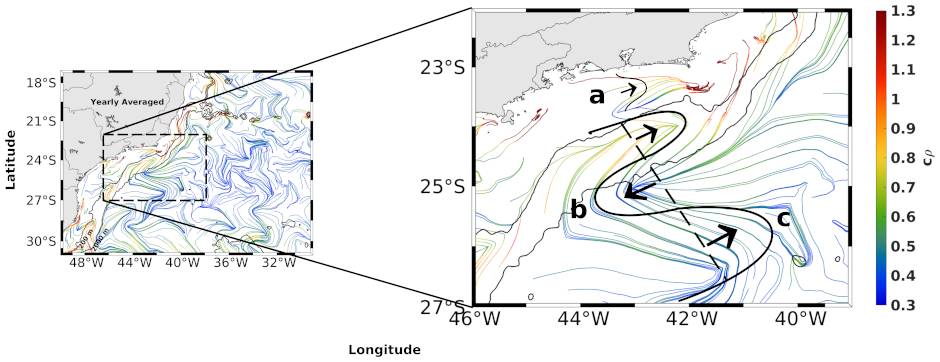}
    \caption{Schematic representation of persistent Lagrangian transport based on the annual averaged cLCSs, colored according to annual $c\rho$. The map to the left shows the full domain of the study and the map to the right is a zoom in to the dashed ellipse. The 200 and 2000 $m$ depth contours are represented by thin black lines.}
    \label{fig:06:transport}
\end{figure}

The BC axis (see MKE in Fig.~\ref{fig:02:eke}c and~\ref{fig:02:eke}d, and SST and cLCSs chevrons in Fig.~\ref{fig:01:mur}) is positioned between anticyclonic and cyclonic structures (Fig.~\ref{fig:05:ssh}e). 
As it flows from northeast to southwest, the eastern flank of the BC gains counterclockwise rotation offshore of the 2000 $m$ isobath. 
The alongshore flow in the east-west oriented shelf near region 7 (Fig.3), becomes offshore flow from Cabo Frio as it approaches the 200 $m$ isobath, and is fed by a clockwise circulation over the 200 $m$ isobath, connecting with the counterclockwise circulation offshore. 
Thus, the monthly-mean Eulerian velocity suggests limited cross-shelf transport between the 200 and 2000 $m$ isobaths (Fig.~\ref{fig:05:ssh}e), that coincides with weak attraction $c\rho$ (Fig.~\ref{fig:03:attraction}b and~\ref{fig:05:ssh}), and coincides also with the Eulerian-mean anticyclonic and cyclonic structures (b and c in the Fig.~\ref{fig:06:transport}) found on both sides of the BC axis. 

Based on this, we propose a schematic representation of the persistent meandering between 23 and 27S (Fig.~\ref{fig:06:transport}). 
The strength of normal attraction along the cLCSs deformed as chevrons by the BC is indicative of the kinematics: along the BC core between 25--26S, cLCSs are weak ($c\rho$ ~ 0.3 in logarithmic scale) indicating the core of the BC, where normal attraction is weak. 
Between 24--25S however, the BC core has higher values of $c\rho$ (~0.6  in logarithmic scale) reflecting an increase in normal attraction that is associated with the cross-shelf circulation described above. 
Between 23--24S, cLCSs reach high $c\rho$ values (>0.8  in logarithmic scale) reflecting the coastal tributary to the alongslope flow, that is related to the coastal upwelling jet off Cabo Frio (see a in the Fig.~\ref{fig:06:transport}).
There are similar dipoles adjacent to the BC at different latitudes (Fig.~\ref{fig:05:ssh}d, e, f), and the cLCSs and $c\rho$ show a similar response (Fig. S4).

\subsection*{Variability of surface flow assessed through drifters}

\begin{figure} 
    \centering
    \includegraphics[width=0.9\textwidth]{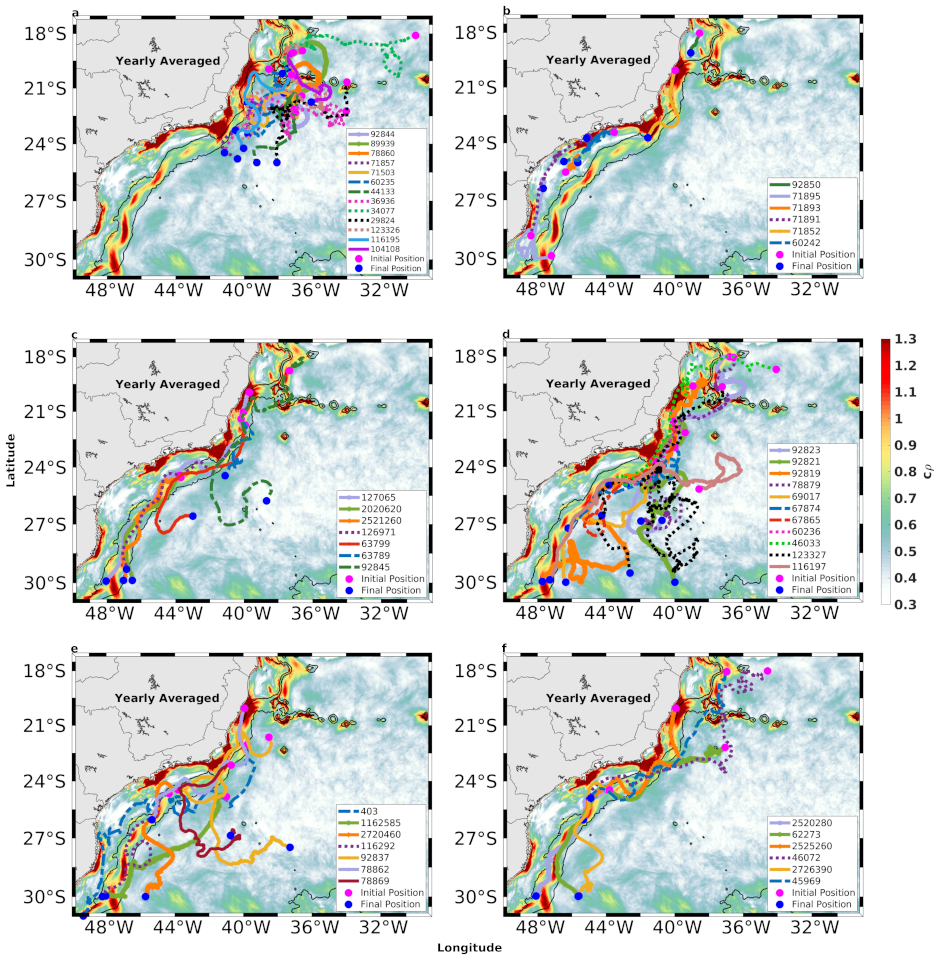}
    \caption{Selected trajectories of satellite-tracked drifters with drogues of 15 $m$ depth, interpolated at 6
    hour intervals, distributed by NOAA'S Global Drifter Program, and of iSPHERE, distributed by Prooceano with consent of PetroRio, superposed to the annual-mean $c\rho$.(a) drifters trapped in eddies and meanders near the Vitória-Trindade Seamount Chain and Almirante Saldanha seamount; (b)drifters confined in weakly attracting regions or inshore of the 200 $m$ isobath; (c) drifters traveled inshore of the BC core, between the 200 and 2000 $m$ isobaths; (d) drifters moved offshore of the 2000 $m$ isobath; (e) drifters moved onshore-offshore-onshore; (f) drifters moved onshore-offshore-onshore. Depth contours of 200 and 2000 $m$ are represented by thin black lines. The ID of each drifter is shown in the insert of each frame.}
    \label{fig:07:drifters}
\end{figure}

From a total of 352 trajectories of satellite-tracked 15 $m$ drogued drifters, distributed by NOAA's Global Drifter Program (GDP), and iSPHERE, distributed by Prooceano, with consent of PetroRio, interpolated to 6-h intervals, a minority crossed at some time the 2000 $m$ isobath and less than 30\% of drifters cross the 200 $m$ isobath (Table~\ref{tab:drifters}). 
Some of these trajectories are plotted with the annual $c\rho$ maps (Fig.~\ref{fig:07:drifters}) to highlight the quasi-steady Lagrangian transport patterns associated with the BC around 2000 $m$ isobath.
The observed transport patterns show a region of flow with a number of drifters spending several days trapped in eddies and meanders near the Vitória-Trindade Seamount Chain (Fig.~\ref{fig:07:drifters}a, see also Region 2 in Fig.~\ref{fig:03:attraction}a) and Almirante Saldanha seamount (Region 5 in Fig.~\ref{fig:03:attraction}a). 
Drifters may also spend some time confined in weakly attracting regions, represented by low $c\rho$ values or confined inshore of the 200 $m$ isobath which also has low $c\rho$ values except at some locations near the coastline (Fig.~\ref{fig:07:drifters}b).

\begin{table} 
    \centering
    \begin{tabular}{|l|l|}
    \hline
    \textbf{Regions}            & \textbf{Drifters (\%)}*   \\ \hline
    weak attraction region 01   & 10.51                     \\ \hline
    weak attraction region 02   & 29.54                     \\ \hline
    weak attraction region 03   & 16.47                     \\ \hline
    weak attraction region 04   & 10.51                     \\ \hline
    weak attraction region 05   & 17.04                     \\ \hline
    \end{tabular}
    \caption{Percentage of satellite-tracked drifters with 15 $m$ drogues distributed by NOAA'S GDP, with trajectories interpolated in 6 hour intervals, that crossed the 2000 $m$ isobath along a region of weak attraction, as indicated in Fig.~\ref{fig:03:attraction}b. The same drifter can be counted more than once if it crossed more than one region of weak attraction. *The percentage is related to the total of 352 drifters.}
    \label{tab:drifters}
\end{table}

When drifters travel just inshore of the BC core, between the 200 and 2000 $m$ isobaths, there is some meandering (Fig.~\ref{fig:07:drifters}c); drifters that move offshore of the 2000 $m$ isobath do so where there are persistent cyclonic structures (Fig.~\ref{fig:05:ssh}e and~\ref{fig:05:ssh}f). 
Further examples of cross-shore transport show that the persistent cyclonic structures between 23--25S, 26.5S and 29S (Fig.~\ref{fig:05:ssh}e and~\ref{fig:05:ssh}f) tend to influence where drifters will move onshore or offshore (Fig.~\ref{fig:07:drifters}c--f).
Onshore transport is less likely to cross the 200 $m$ isobath than the 2000 $m$ isobath.
The cross-shelf flow of drifters between 24S and 27S (Fig.~\ref{fig:07:drifters}c--e) is more likely over regions of low $c\rho$ than over regions of high $c\rho$.

The effect of persistent meandering and eddy-like structures (Fig.~\ref{fig:05:ssh}a and\ref{fig:05:ssh}b) on satellite-tracked drifters can clearly be seen as high values of Probability Density Estimates (PDE, see Methods section) measuring the likelihood of drifters visiting a region. 
In February and May, regions of high PDE tend to be confined within $c\rho$ maxima that occur close to upwelling regions and along steep bathymetric features (Fig.~\ref{fig:03:attraction}, regions 2 and 5).

\begin{figure} 
    \centering
    \includegraphics[width=0.9\textwidth]{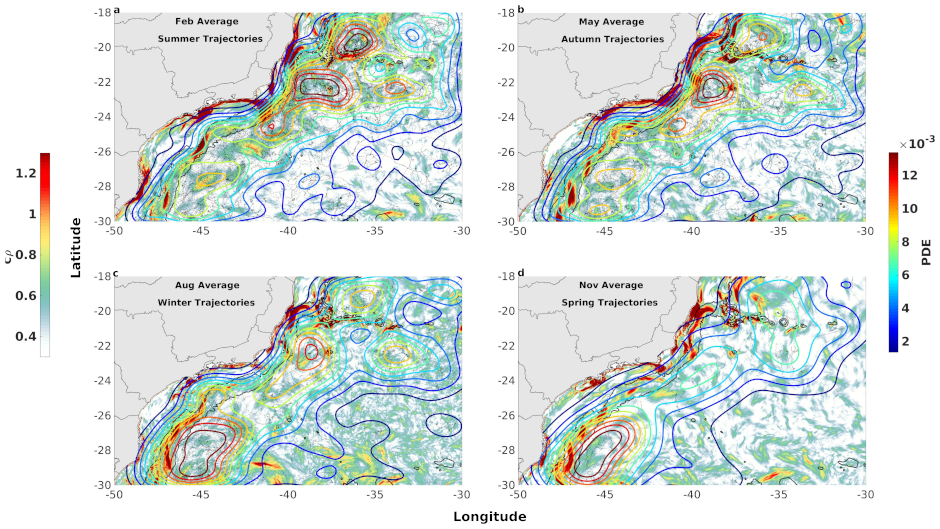}
    \caption{The PDE (coloured contours) of (a) all austral Summer trajectories (Jan-Feb-Mar) over the $c\rho$ distribution for February (color shading), (b) all austral Autumn trajectories (Apr-May-Jun) over the $c\rho$ values for May (color shading), (c) all austral Winter trajectories (Jul-Aug-Sep) over the $c\rho$ values for August, and (d) all austral Spring trajectories (Oct-Nov-Dec) over the $c\rho$ distribution for November. Depth contours of 200 and 2000 $m$ are represented by black lines.}
    \label{fig:08:pde_crho}
\end{figure}

Between 24--30S, PDE values tend to be well aligned with the slope, with values diminishing considerably towards the coast at any given latitude.  
Exceptions are localized near 25S in February, May and August, when medium values of PDE ($\approx6-10\times10^{-3}$) can be seen just inshore of the 200 $m$ isobath. 
PDE maxima ($>11\times10^{-3}$) are confined offshore of, and adjacent to the 2000 $m$ isobath. 
All drifters can be seen in the supplemental information (see Supplementary Fig. S6). 
The drifter PDE suggests seasonal variability with PDE maxima ($>11\times10^{-3}$) located only north of 24S in the summer and autumn (Fig.~\ref{fig:08:pde_crho}a and~\ref{fig:08:pde_crho}b), with winter being a transition as PDE maxima diminish north of 24S and increase south of 26S (Fig.~\ref{fig:08:pde_crho}c), and in spring PDE maxima ($>11\times10^{-3}$) is limited to south of 26S. 

In all cases PDE maxima tends to be just offshore of the 2000 $m$ isobath, the only exception being a maximum centered near 22S and 34W, mainly found in summer, autumn and winter. 

The isolated region next to the coastline between 24 and 27S has a larger zero PDE region in autumn and winter when $c\rho$ is also negligible in this region (Fig.~\ref{fig:08:pde_crho}b and~\ref{fig:08:pde_crho}c) relative to $c\rho$ in summer and spring (Fig.~\ref{fig:08:pde_crho}a and~\ref{fig:08:pde_crho}d). 
In all seasons this coastal region is well isolated. 
Another region that tends to be isolated is the coastline north of 20S, especially in autumn, winter and spring (Fig.~\ref{fig:08:pde_crho}b--d).

\subsection*{Oil spill at Frade's Field}

The cLCSs and $c\rho$ computations have been successfully used to estimate the likely trajectory of oil spills such as during the accidents in the Gulf of Mexico in 1979 and in 2010\cite{duran2018extracting}.

\begin{figure} 
    \centering
   \includegraphics[width=0.9\textwidth]{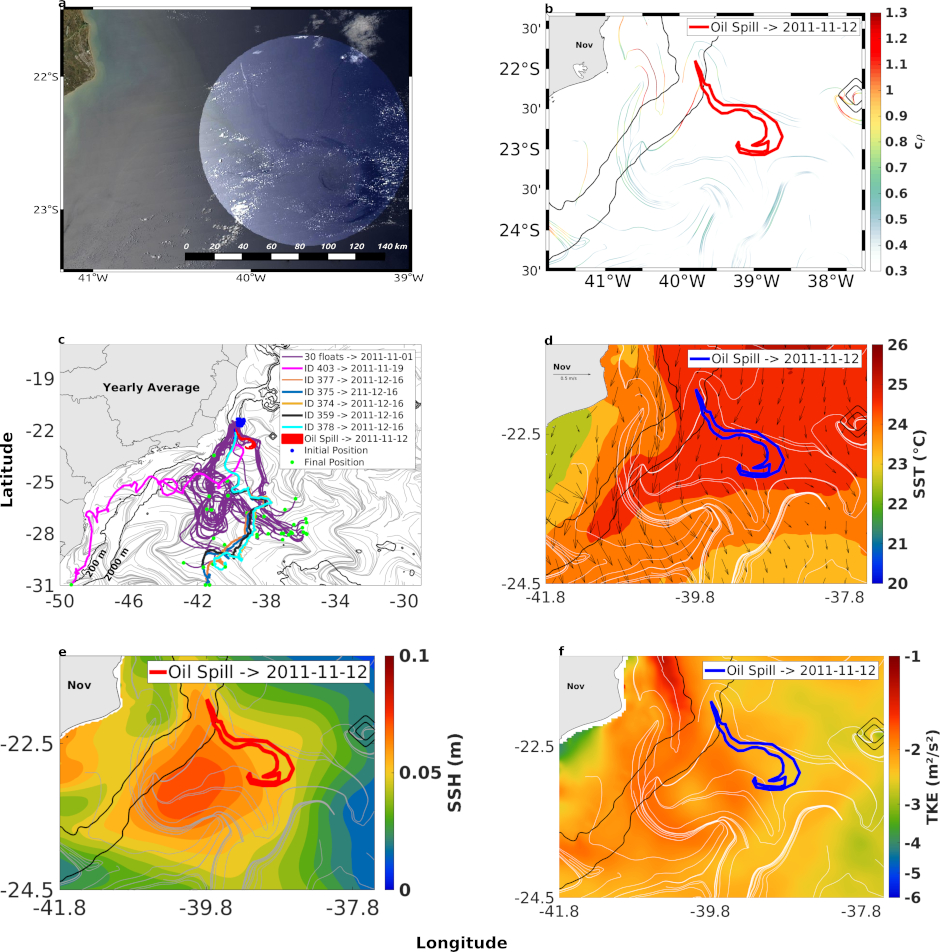}
    \caption{(a) The Chevron oil spill as observed by a true-color MODIS satellite image composite (courtesy of Petrobras) from 12th November, 2011 at 10:30 (UTC--3) in the Frade’s field. (b) The shape of Chevron’s oil spill (red polygon) over the cLCSs for November coloured according to their strength of attraction $c\rho$. (c) The chevron’s oil spill (red filled polygon), the 60 day trajectories of 30 ROMS synthetic-drifters launched in November 1st, 2011 (purple lines), and the trajectories of 06 iSphere drifters (courtesy of Prooceano and PetroRio S.A.) launched between November and December of 2011 (magenta, orange, blue, yellow, gray and cyan lines) plotted over the annual climatological squeezelines (grey contours). The dates of deployment and ID of each iSphere are identified in the figure insert. (d) Chevron oil spill (blue polygon) over the ROMS surface velocity (black arrows), MUR SST (color shading), and cLCSs (white lines), all averaged for November. (e) Chevron’s oil spill (red polygon) over the mean ROMS SSH (color shading) and cLCSs (white lines) for November. (f) Chevron’s oil spill (blue polygon) over the ROMS TKE (color shading, in log scale) and the November cLCSs (white lines). Depth contours of 200 and 2000 $m$ are represented by thin black lines.}
    \label{fig:09:frade}
\end{figure}

We show that performing the same computations using a free-run eddy-resolving ocean model it is possible to achieve a good level of agreement between cLCSs and the behavior of a much smaller offshore oil spill that occurred at the Frade's Field (Fig.~\ref{fig:09:frade}a), located 120 $km$ off the coast of Rio de Janeiro State (Brazil), on November 7, 2011.
The spill reached 160 $km^2$ in less than 15 days, being contained on December, 30 of the same year (Fig.~\ref{fig:09:frade}b).

The agreement with cLCSs can be qualitatively assessed by comparing the oil spill trajectory with the trajectories of 30 synthetic drifters released in the same month of the accident and with six satellite-tracked iSphere drifters deployed between November and December, 2011 (Fig.~\ref{fig:09:frade}c).
The synthetic drifters were allowed to move for 60 days, which is the same period the real spill progressed before its final containment.

Part of the synthetic drifters (purple trajectories to northwest in Fig.~\ref{fig:09:frade}c) agree with the observed spill trajectory (red shape) and with one of the iSphere drifters deployed in November (magenta trajectory -- ID 403 -- in Fig.~\ref{fig:09:frade}c), for the most part moving along a cLCSs. 
The other part of the synthetic floats (purple trajectories to southwest in Fig.~\ref{fig:09:frade}c) correspond with iSphere trajectories (orange, blue, yellow, gray and cyan lines in Fig.~\ref{fig:09:frade}c) deployed in December, and the only other cLCSs originating where the oil spill originated and where synthetic and real drifters were released.
Thus, synthetic drifters, iSphere trajectories and cLCSs are all in good agreement: there are two main transport patterns originating at the spill location.  
While the oil spread along one of the cLCSs (Fig.~\ref{fig:09:frade}b), it did not follow the monthly mean surface currents for November (Fig.~\ref{fig:09:frade}d). 
Indeed the mean Eulerian velocity is often perpendicular to cLCSs originating at the location of the spill. 

Notice that cLCSs are accurate indicators of the first part of the drifter trajectories -- they were designed to extract likely transport patterns over periods of about one week. 
The oil spill started in November near an upwelling region (Region 6 in Fig.~\ref{fig:03:attraction}b, see also Fig.~\ref{fig:09:frade}d) and permeated the southeast near the mean position of a cyclonic feature (Fig.~\ref{fig:09:frade}e and~\ref{fig:09:frade}f, see also~\ref{fig:05:ssh}d). 
The oil spill final positions coincide with the persistent squeezelines deforming into chevrons in November (23S and 39W) and advecting the oil spill away from the 2000 $m$ isobath. We note how the cLCSs agree with the thermal fronts (Fig.~\ref{fig:09:frade}d), as with the low-frequency SSH (Fig.~\ref{fig:09:frade}e) and EKE (Fig.~\ref{fig:09:frade}f) distribution.

\subsection*{A large-scale oil contamination in Northeast Brazil with unknown origin}

In 2019, Brazil experienced an oil-related environmental emergency that impacted a large number of beaches as far south as the state of Rio de Janeiro, reaching an extension of almost 4000 $km$ \cite{ibama2019info7,marinha2019manchas}.

Since late August, when it was first detected, more than a thousand beaches have reported occurrences of oil patches, including 12 marine protected areas \cite{ibama2019info7}. By November, over 2000 metric tons of oil were removed from these beaches \cite{guardian2019manchas} and there is still no indication or evidence of its origin \cite{ibama2019info7}.
Crude oil may drift as shallow subsurface patches, making it difficult to use satellite sensors for monitoring. 

One of the most important Marine Protected Areas, the Abrolhos Bank, was oiled in early November 2019.
This area is internationally recognized as a marine biodiversity heritage \cite{herrero2019abrolhos,wwf2019abrolhos,ci2019,safety4sea2019abrolhos} and has been included in the 16th round of bidding for exploration and production of oil and natural gas, under the concession regime, opened in December 2018 (resolution CNPE n\degree17/2018 and CNPE n\degree03/2019; \cite{anp2019}).

\begin{figure} 
    \centering
    \includegraphics[width=0.9\textwidth]{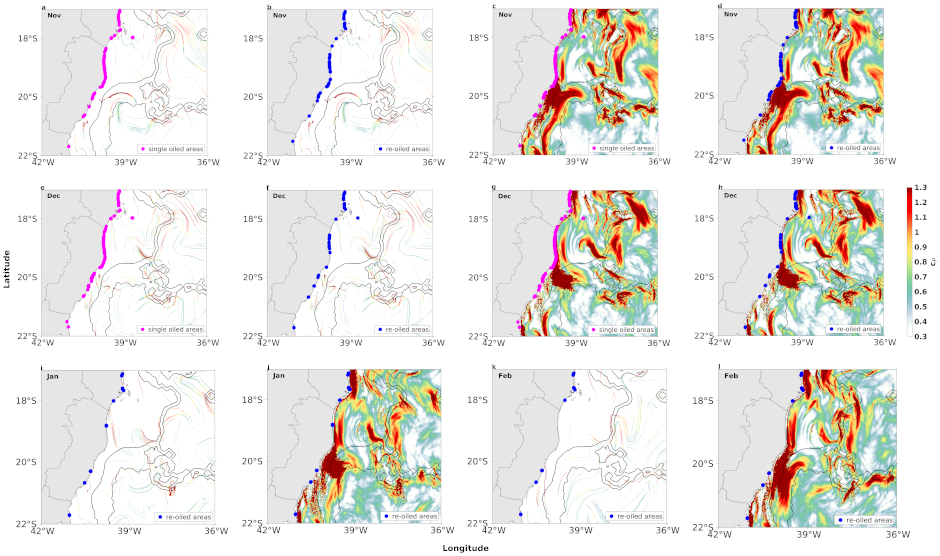}
    \caption{Location of beaches along the Brazilian coast within the study area that were oiled for the first time (pink dots) and that were re-oiled (blue dots) in November (a-d) and December (e-h) of 2019, and January (i-j) and February (k-l) of 2020. Depth contours of 200 and 2000 $m$ are represented by thin black lines. Oiled sites are superposed to monthly mean cLCSs (colored lines) and  $c\rho$ (color shading) to indicate transport patterns and particle attraction sites.}
    \label{fig:10:northeast}
\end{figure}

We show the location of known oiled beaches located within our study area as per November of 2019 (Fig.~\ref{fig:10:northeast}a--d), December of 2019 (Fig.~\ref{fig:10:northeast}e--h), January of 2020 (Fig.~\ref{fig:10:northeast}i--j), and February of 2020 (Fig.~\ref{fig:10:northeast}k--l)  (IBAMA, 2020). 

In this area, the oil beached for the first time during the months of November (Fig.~\ref{fig:10:northeast}a--c) and December (Fig.~\ref{fig:10:northeast} and 10 $g$).
The floating oil moved from north to south following the BC and, although most of the contaminated coastline is outside our model domain, it is possible to note that regions of maximum values of $c\rho$ ($>$ 1.3, 3.7 in linear units) are strong indicators of oil beaching, both for a single time (Fig.~\ref{fig:10:northeast}a, c, e and g) and for a repeated occurrence (Fig.~\ref{fig:10:northeast}b, d, f, h--l).

%% file: sections/02_discussion.tex
We offer, for the first time, an integrated representation for the role of eddies and meanders in shaping the mean flow of the BC based on the calculation of cLCSs.
The quasi-steady nature of these Lagrangian structures allows the identification of the pervasive and consistent influence of mesoscale features in this western boundary current.
Climatological squeezelines deforming into chevron shapes can be seen along the axis of the mean BC coinciding well with chevron shapes from satellite SST, as it is advected by the mean flow.
These structures characterize the BC core at the surface positioned along the 2000 $m$ isobath, with a good seasonal agreement between model and high resolution satellite data (Fig.~\ref{fig:01:mur}). 
High resolution model output indicates that cLCSs deformation also responds to zonally persistent cross-shelf SSH transition from positive (high) values near coastline to low (negative) values between 200 and 2000 $m$ and back to positive (high) offshore from the 2000 $m$ isobath (Fig.~\ref{fig:05:ssh}a and~\ref{fig:05:ssh}b).
Zonally-paired cyclonic and anticyclonic structures are embedded in this transition, also causing the cLCSs to deform into chevrons.
In a recent work, \cite{beron2020stability} found similar chevrons straddling The Malvinas Current lagrangian axis, and showed it was a persistent structure by superimposing shearless-parabolic LCSs that behaves as as a cross-shelf transport barrier.
Indeed, we show that hyperbolic cLCSs can also identify the deformation of fluid as chevrons along the core of the BC, in close agreement with monthly-mean satellite SST chevron-like advection patterns (Fig.~\ref{fig:01:mur}). 

A number of independent studies have discussed the presence of mesoscale features along the BC \cite{schmid1995vitoria,campos1995water,silveira2004baroclinic,silveira2008meander, calado2006parametric,chen2019seasonal}, frequently invoking the interaction with sharp topographic gradients and surface-subsurface flow shear as inducing the formation of cyclonic and anticyclonic meanders with consequent pinching off of eddies \cite{arruda2013events,soutelino2013roles}.
During the summer, the cLCSs deformation represented by the chevrons covers a large area with seamounts and sharp change in coastline orientation, coincident with EKE maxima (-2 $m^2/s^2$, in logarithmic scale, 0.13 $m^2/s^2$ in linear units) extending offshore of the Abrolhos Bank and between 21S and 26S (Fig.~\ref{fig:02:eke}a).
During the winter, EKE maxima is less organized (Fig.~\ref{fig:02:eke}b), possibly as a result of the increase in the number of northward-moving cold atmospheric fronts.
The presence of persistent meanders and eddies along the BC axis is also evident in the seasonal (monthly means for March and August) EKE and MKE maps (Fig.~\ref{fig:02:eke}a--d), monthly-model SSH (Fig.~\ref{fig:05:ssh}), seasonal PDE of drifter trajectories (Fig.~\ref{fig:08:pde_crho}) and model monthly mean surface velocities (Fig. S5).
The orientation and location of cLCSs tend to follow these known features, with our model SSH indicating both cyclonic (Vitória Eddy, Fig.~\ref{fig:05:ssh}c) and anticyclonic rotation (Abrolhos Eddy, Fig.~\ref{fig:05:ssh}c), near the Abrolhos region and Vitória-Trindade chain respectively (locations 1, 4, and 2 in Fig.~\ref{fig:03:attraction}a). 
Intense flow shear caused by the Abrolhos Bank eastward shelf projection and the Vitória-Trindade seamount chain are both evident as high $c\rho$ values ($>$ 1, 2.7 in linear units, see Fig.~\ref{fig:04:crho}) and chaotic trajectories of drogued drifters (Fig.~\ref{fig:07:drifters}a).
As the BC moves southwards, it crosses a wind-induced upwelling region around 21S \cite{castelao2006upwelling,rodrigues2001numerical} in the Cabo de São Tomé Region (location 6 in Fig.~\ref{fig:03:attraction}a, see also Fig.~\ref{fig:05:ssh}d and S8a), characterized by persistent year-round high $c\rho$ (Fig.~\ref{fig:04:crho}). 
The cLCSs deformation in the São Tomé eddies depicted in Fig.~\ref{fig:05:ssh}d has a similar shape to those found along the BC axis further south, but with the offshore eddy displaying a cusp bent towards the direction of anticyclonic rotation.
As these chevrons are formed distant from the BC axis, their shapes tend to conform to the eddy rotation.
The onshore eddy is a semi permanent cyclonic feature (see Fig. S8a--c) originated from the detachment of a BC unstable meander likely induced by baroclinic instability \cite{calado2008feature,campos2006equatorward,lima2016assessment}. 
The persistent meanders and the cyclonic eddies they generate are known to induce or enhance coastal upwelling \cite{calado2010eddy}, here we associate the upwelling with high attraction values ($>$ 1, 2.7 in linear units) of $c\rho$ (Fig.~\ref{fig:04:crho}).
There is a Lagrangian confluence region just south of the upwelling of Cabo Frio (location 7 in Fig.~\ref{fig:03:attraction}a) associated with the presence of cyclonic-anticyclonic features on both sides of the 2000 $m$ isobath near 25S (Fig.~\ref{fig:05:ssh}e, Fig. S8a--c).
Here, our computations of $c\rho$ and cLCSs accurately captured the main elements of surface circulation known as a current-eddy-upwelling region \cite{calado2010eddy}, offering a spatio-temporally Lagrangian integrated view of a dynamically complex system.
Further south, Cabo de Santa Marta (location 9 in Fig.~\ref{fig:03:attraction}a) is dynamically similar to Cabo Frio with alternating anticyclonic (offshore) and cyclonic (inshore) flows (Fig.~\ref{fig:05:ssh}f) that coincide with satellite SST advection (see Fig. S8d) enhanced by baroclinic instability \cite{silveira2008meander}.
The persistent shelf-break upwelling results from the interaction of the BC with coastline orientation and shelf topography \cite{palma2009disentangling}.
During the summer, wind forcing enhances upwelling and in the winter a northward flow advects cold, low-salinity water from the Plata river \cite{campos2013seasonal} (Fig.~\ref{fig:01:mur}a).
The zonally organized cLCSs around Cabo de Santa Marta have chevrons structures over the MKE maxima (Fig.~\ref{fig:02:eke}c and~\ref{fig:02:eke}d) along the axis of the mean BC flow. 

The cLCSs extracted quasi-steady Lagrangian transport patterns associated with persistent meanders and rotational structures, as indicated in different regions (Fig.~\ref{fig:05:ssh}c--f).
When the BC axis is over the 2000 $m$ isobath, south of about 22S, frontal meanders develop as a response to baroclinic instability due to vertical shear associated with the BC and the Intermediate Western Boundary Current flowing below 500 $m$ depth \cite{silveira2008meander}.
As these meanders grow they can cause a reversion in the surface flow in the inshore front of the BC.
We have used the annual average cLCSs to propose an schematic representation, shown in Fig.~\ref{fig:06:transport}, of the quasi-stationary persistent meander flow based on the example of Fig.~\ref{fig:05:ssh}e.
A similar baroclinic flow was obtained by \cite{calado2010eddy} with an ocean model experiment with an horizontal resolution of 13 $km$ and 20 $\sigma$ levels showing the existence of a quasi-standing vorticity wave pattern in the region.
The typical cyclonic meandering between 200 and 2000 $m$ depth is also known to induce shelf-break austral Summer upwelling \cite{campos2000shelf}.
\cite{campos1996experiment} described the existence of pairs of eddies with opposite rotations south of Cabo Frio in the same way as our schematic representation, but were unable to offer an explanation for that. 
This persistent feature associated with high attraction $c\rho$ values ($>$ 1, 2.7 in linear units) probably contributes for frequent revisit of drifters during the austral summer and winter (Fig.~\ref{fig:08:pde_crho}a and~\ref{fig:08:pde_crho}b).

Interpreting cLCSs is not always a simple task so that the integration of different and independent data is needed.
For that purpose, Lagrangian drifters and synthetic floats released at specific sites have been successfully used to assess the significance of the computed cLCSs \cite{duran2018extracting,gough2019persistent}.
Despite the seasonal variability observed in the satellite SST and model-derived EKE and cLCSs, the BC flow tends to act as a transport barrier for particles.
The computed PDE for 352 satellite-tracked drifters shows that they mostly concentrate in large patches along the 2000 $m$ isobath (Fig.~\ref{fig:08:pde_crho}, see also Supplementary Fig. S6 for a plot of all trajectories).
There is a clear spatio-temporal pattern in the distribution of trajectories, from February to May they concentrate at the northern section equatorward of 24S and to the south between August and November. 
The prevalence of high attraction $c\rho$ values ($>$ 1, 2.7 in linear units) also follow a similar trend as depicted from the monthly climatological attraction strength maps in Fig.~\ref{fig:04:crho}.  

The strong variability in this alongshelf current is suggested by the presence of intense eddy activity (Fig.~\ref{fig:02:eke}a and~\ref{fig:02:eke}b, Fig.~\ref{fig:05:ssh}d--f), with the magnitude of EKE maxima being comparable to the MKE maxima (Fig.~\ref{fig:02:eke}c,d). 
However, MKE is centered along the 2000 $m$ isobath, while EKE maxima is centered just inshore, between the 200 and 2000 $m$ isobaths. 
The location of TKE maxima is similar to the location EKE maxima with contributions from the MKE maxima, again suggesting the persistence of eddy activity (Fig.~\ref{fig:02:eke}e,f).
The deformation of climatological squeezelines crossing regions with high and low mean model SSH and aligning with surface Eulerian velocities (Fig. S5) show a consistent signature.
This is most evident poleward of about 23S, where SSH gradient of low SSH between 200 and 2000 $m$ isobaths, and high SSH offshore of the 2000 $m$ isobath (Fig.~\ref{fig:05:ssh}a and~\ref{fig:05:ssh}b), seem to concentrate most of the flow horizontal shear.
Nearly 30\% of the 352 drogued drifters used in this study (Table~\ref{tab:drifters}) managed to cross regions of weak attraction (Fig.~\ref{fig:03:attraction}b, see also Fig.~\ref{fig:07:drifters}c drifter 127065; \ref{fig:07:drifters}e drifters 403, 78862, 2720460; ~\ref{fig:07:drifters}f drifters 62273, 2525260).
In fact, the PDE of drogued drifter positions (Fig.~\ref{fig:08:pde_crho}) suggest that high $c\rho$ over the slope behave as a transport barrier around 200 $m$ isobath, the likelihood of a drifter crossing this threshold is about an order of magnitude smaller than the likelihood of not crossing.
When drifters cross the 2000 $m$ isobath towards the 200 $m$ isobath, they tend to do so along a section with low $c\rho$. 
Similarly, drifters released inshore of the 200 $m$ isobath tend to be confined to the shelf (Fig.~\ref{fig:07:drifters}b). 

Our study domain covers two of the most important oil-producing areas of the southwest Atlantic, the Santos and Campos basins, that are responsible for 87\% of total oil production in Brazil.
Our results highlight the complex nature of surface transport along the BC and the challenges it poses to those involved in modeling oil spill trajectory as part of contingency planning and emergency response.
Using examples of two different oil spill events, we show that cLCSs and $c\rho$ computations provide new information that is relevant for a detailed assessment of surface transport organization. 
Our results for the Frade’s spill agree with offshore transport that can be seen with drifting buoys (Fig.~\ref{fig:09:frade}c), monthly SST-satellite (Fig.~\ref{fig:09:frade}d), and the maxima EKE (Fig.~\ref{fig:09:frade}f) offshore from the 2000 $m$ isobath. 
We can see from (Fig.~\ref{fig:09:frade}c that only one iSphere drifter (ID 403) and some of the synthetic drifters managed to reproduce the movement of the oil spill with the same accuracy as the transport determined by the cLCSs (Fig.~\ref{fig:09:frade}b).
The usefulness of cLCSs for oil spill planning and response is again demonstrated here.
At the Frade's field oil spill location, the synthetic drifters and iSphere trajectories show two different transport patterns, with the oil following one of them. 
The fact that both transport patterns are clearly depicted by the cLCSs that originate at the spill's origin further support the potential of using these Lagrangian structures to constrain the most likely oil trajectories during an emergency response.
By comparing to transport patterns, it is shown here once again that the time-averaged Eulerian velocity can be misleading the Eulerian velocity tends to be perpendicular to the simulated and observed Lagrangian transport patterns, which are accurately depicted by cLCSs (Fig.~\ref{fig:09:frade}d, see additional examples in the supplementary information of \cite{duran2018extracting}).
In this study, we show that interpreting cLCSs is not always straightforward.
The interpretation, and consequently the identification of dominant transport patterns, is supported by comparisons between cLCSs with time-mean Eulerian fields such as SST, SSH and TKE (Fig.~\ref{fig:01:mur},~\ref{fig:02:eke} and~\ref{fig:05:ssh}), and with drifters (e.g. Fig.~\ref{fig:08:pde_crho} and~\ref{fig:09:frade}).

Recently, a large-scale accident oiled nearly 4.000 km of beaches in Brazil between November 2019 and February 2020, for which the origin has not been determined so far.
This gave us the unique opportunity to evaluate how the computed $c\rho$ and cLCSs would have contributed to the observed oil beaching patterns, without consideration to the source of contamination. 
By construction, cLCSs were designed to work for generic oil spills. 
The sequence of reported oiled sites by locals and the Brazilian government suggest that the oil spill should have originated close to the South Equatorial Current bifurcation centered around 10S and 14S \cite{rodrigues2007seasonal}.
Most of the oil dispersed as subsurface patches, yet we found good agreement between the regions of maximum values of $c\rho$ and persistent cLCSs and the first and re-oiled areas in the beaches Brazilian spill (Fig.~\ref{fig:10:northeast}), suggesting which regions are most vulnerable.
Comparing the sites impacted only once, and those that were re-oiled in Fig.~\ref{fig:10:northeast} (pink and blue dots, respectively), clearly the latter tend to happen closer to $c\rho$ maxima ($>$ 1.3, 3.6 in linear units).
The combined use of persistent $c\rho$ and cLCSs with model and observational data (satellite imagery and drifting buoys) showed to be a promising tool to indicate likely oil spill trajectories and beaching sites.

%% file: sections/03_conclusions.tex
We show that by combining cLCSs and $c\rho$ with SST-satellite data, model Eulerian surface velocities, mean SSH, TKE, MKE and EKE, Lagrangian drifters and synthetic drifters, it is possible to gain new insights on how surface ocean transport is organized in a complex weak WBC setting. 
The quasi-steady Lagrangian transport patterns in this western boundary current elegantly captured the role of persistent and recurrent eddies and meandering on the surface transport. 
This novel approach produced consistent results, making it possible to create an integrated representation for the role of mesoscale activity in shaping the mean flow of the BC. 
Accurate representation of surface flows in current systems dominated by instabilities and intense mesoscale activity is particularly challenging, e.g. reconciling Eulerian and Lagrangian views. 
So far, published results in the BC has provided evidences that the interaction of surface and pycnocline-level flows, together with complex bottom topography and sharp changes in the coastline orientation produce a number of persistent mesoscale features \cite{campos1995water,campos1996experiment,silveira2008meander}.
The time-mean Eulerian flow may not be representative of material transport, making it difficult to accurately describe at material transport the surface of the ocean. 
We overcome this limitation by describing the surface flow of the BC from a Lagrangian point of view, and then connecting it to Eulerian fields such as SSH, MKE and EKE. 
The significance of the above proposed scheme was assessed using two different oil spill events and proved to generate consistent results when compared to the observed spill trajectory and oil beaching.

%% file: sections/04_methods.tex
Our domain is bounded to the north by the Abrolhos National Bank and Vitória-Trindade Seamounts and to the south by the southern limit of Cabo de Santa Marta, between 17--31S and 29--50W.

\subsection*{ROMS velocity data}

We use daily averaged outputs from a ROMS simulation \cite{shchepetkin2005regional,shchepetkin2009correction} with a horizontal resolution of $1/36\degree$ ($\approx$3 $km$)  and 40 terrain-following vertical levels. 
The model simulation was forced every 6 $h$ by the atmospheric fields obtained from the Climate Forecast System Reanalysis (CFSR) and Climate Forecast System (CFSv2) with $\approx$38 $km$ horizontal resolution \cite{saha2010ncep,saha2012ncep,saha2014ncep} and every 5 days lateral open boundary conditions by Simple Ocean Data Assimilation (SODA, version 3.3.1) with horizontal resolution of $0.25\degree$ and 50 vertical levels \cite{carton2018soda3}.
The simulation included the inputs of two rivers (Doce and Paraíba do Sul), using the monthly runoff climatology estimated by Brazilian National Water Agency \cite{ana2018} and river temperature by the Operational Sea Surface Temperature and Sea Ice Analysis (OSTIA) \cite{donlon2012operational}.
Tidal forcing included 08 main tidal constituents, two long periods constituents and three nonlinear harmonic constituents, extracted from the Oregon State University TOPEX/Poseidon Global Inverse Solution -- TPXO version 8 \cite{egbert2002efficient}.
Our free-running simulation was integrated from January 1, 2000 to December 31, 2015, totalling 15 years of experiment, the first 3 years were discarded as spin-up, in order to use the period of model integration in which the surface energy oscillates almost periodically around a steady state \cite{marchesiello2003equilibrium}.

\subsection*{Lagrangian Simulations}

The ROMS simulations contains a built-in float algorithm that allows online tracking of passive floats across the model domain.
Particle trajectories are calculated from the Eulerian velocity fields at each baroclinic time step using the fourth-order Milne predictor and the fourth-order Hamming corrector \cite{narvaez2012modeling,van2018lagrangian}.
Particle simulations were performed to cover two objectives: i) analyze the variability behind the low-frequency Lagrangian transport patterns extracted through cLCSs, and also test the information extracted through cLCSs, like locations of enhanced cross-shelf transport or isolated regions, and ii) reproduce the Lagrangian transport pattern that occurred during the Frade's Field oil spill. 

For the first objective, 30 floats were launched at 28 different points in the study domain.
All 28 launches, with 30 floats each, were carried out in the austral summer and winter at the surface and include a random walk component.
In the austral summer, the floats were launched on the 1st of December 2013 and traveled freely, in the horizontal direction, until the 28th of February 2014.
While in austral winter, the floats were launched on the 1st of June 2006 and traveled freely, in the horizontal direction, until the 30th of September 2006.

For the second objective, we launched, at a single point and once, 30 floats at the sea surface and included a random walk component on 1st of November until 31th of December of 2011.
The simulation coincided with the months and location of Chevron's oil spill in the Frades Field \cite{anp2011}, oil spill data was provided by Petrobras.
During the oil spill six surface drifters (iSpheres) were released by Prooceano, with the permission of PetroRio SA.
The iSpheres \cite{rohrs2012observation,rohrs2015drift} is a low cost, expendable, drifting tracking buoy developed by Metocean Data Systems.
The launch of iSpheres was intended to track and monitor oil spill from Chevron.

\subsection*{Climatological LCS and $c\rho$}

The computation of cLCSs, structures organizing Lagrangian transport, used here is as developed by \cite{duran2018extracting}.
cLCSs are computed using the code in \cite{duran2019ciam}.
The sea-surface velocity data is obtained from daily outputs of a 13-year ROMS simulation. 
The climatology of the superficial velocity was obtained by averaging each day of the time series, defining a 365 day climatology, disregarding, therefore, the leap days.
Further description of the method can be found in \cite{duran2018extracting}, a description of the computations and the code, can be found in \cite{duran2019ciam}.
Trajectories were integrated using a 4th/5th order Runge-Kutta method, with step adaptation, and cubic interpolations. 
The trajectory integration was over 7 days periods ($T = -7$ days), for each initial condition in space ($x_0$) and in time ($t_0$).
An adequate time-scale to extract recurring or persistent transport related to mesoscale structures \cite{duran2018extracting}. 
The computations use a numerical grid of $1024\times878$, with an auxiliary computational grid of 2.03 $km$ to the north, south, east and west of each grid point.

\subsection*{Observed surface trajectories and their Probability Density Estimate (PDE)}

We used 352 satellite-tracked drifters with drogues at 15 $m$ depth distributed by NOAA'S Global Drifter Program -- GDP \cite{elipot2016global}, with trajectories interpolated every 6 hours and spanning 13 years of data to compute a Probability Density Estimate (PDE) of drifter trajectories.
The PDE is calculated using a Probability Density Function, $PDF(\rho,t \mid \rho_0,t_0)$ \cite{murray2011remarks,silverman1986density}, with the initial positions of each trajectory being $\rho_0=(x_0,y_0)$ at time time $t_0$, and the final position $\rho=(x,y)$ at time $t$. And regions with high incidence of trajectories were obtained with a Kernel Density Estimation \cite{epanechnikov1969non} in smoothed, approximately, with $3\degree$ x $3\degree$ boxes at 900 points, equally spaced, calculated according to \cite{epanechnikov1969non,silverman1986density}. 
We adopted a lagrangian time scale of 3-days \cite{assireu2003surface} for each trajectory, based on estimated diffusion coefficient between $6\times10^6$ and $9.1\times10^7 cm^2s{-1}$ and a lagrangian time scale between 1 and 5 days for SWA.
The Lagrangian integral time scale represents the time under which the speed at two different points in time remains autocorrelated, the interval of maximum time that the memory effect on the displacement of the particles is verified in a fluid.

For these trajectories we use a probabilistic approach, using Probability Density Estimate (PDE) of drifter trajectory movements. 
For the statistical analysis the drifter data was separated into three days trajectories based on the typical Lagrangian time scale for this region \cite{schmid1995vitoria,assireu2003surface}.

\subsection*{Auxiliary data}

The sea-surface temperature was obtained from the global daily-SST data of the Multi-scale Ultra-high Resolution (MUR) sensor \cite{chin2017multi}. 
The MUR provides data with spatial resolution of $0.01\degree$, approximately 1 $km$ intervals.

We estimate the distribution of kinetic energy per unit mass for the mean and eddy fields.
The TKE represents the sum of the MKE, the energy of the 
mean circulation and the EKE, the fluctuating part of the absolute velocity \cite{schmid1995vitoria,assireu2003surface}. 
MKE, EKE and TKE were calculated as the Equations~\ref{eq:mke}, \ref{eq:eke} and \ref{eq:tke}:

\begin{equation}
    MKE = \frac{1}{2}\left (\bar{\mathbf{u}}^2+\bar{\mathbf{v}}^2 \right)
    \label{eq:mke}
\end{equation}

\begin{equation}
    EKE = \frac{1}{2}\left (\overline{\mathbf{u}'^2}+\overline{\mathbf{v}'^2}  \right )
    \label{eq:eke}
\end{equation}

\begin{equation}
    MKE = MKE + EKE
    \label{eq:tke}
\end{equation}

Where $\bar{\mathbf{u}}$ and $\bar{\mathbf{v}}$ are the monthly mean surface current velocities computed from the daily means, and $\mathbf{u}'$ and $\mathbf{v}'$ are the departures from the mean.
MKE, EKE and TKE are all in $m^2s^{-2}$.

%% file: sections/10_acknowledgements.tex
This work has been supported by Coordenação de Aperfeiçoamento de Pessoal de Nível Superior - Brasil (CAPES) - Finance Code 001.
The code for cLCS and it’s acknowledgments are available at \url{https://bitbucket.org/rodu/clcss}.
We thank E. van Sebille for the initial help with float releases.
We thank F. J. Beron-Vera for helpful discussions.
We thank Giullian N. L. dos Reis for processing the satellite MODIS data.
We thank C. L. G. Batista, A. S. Ipia for discussions and constructive comments on the text.
We thank L. P. Pezzi of LOA (\url{http://www.dsr.inpe.br/DSR/laboratorios/LOA_OBT.pdf}) for authorization to use the Kerana cluster.
We thank PROCEANO (\url{http://prooceano.com.br/site/}) and PetroRio-S.A (\url{https://petroriosa.com.br/}) for the iSpheres data.
We thank Petrobras (\url{http://www.petrobras.com.br/en/}) for the shape and image of MODIS referring to the oil spill occurred in November 2011.
We thank IBAMA for the oil spill locations data (\url{http://www.ibama.gov.br}).
We thank the developers of Regional Ocean Modeling System (\url{https://www.myroms.org}) and the Jet propulsion Laboratory for providing the Multi-scale Ultra-high Resolution (MUR) Sea Surface Temperature (SST) Analysis which is available at (\url{https://podaac.jpl.nasa.gov/dataset/MUR-JPL-L4-GLOB-v4.1}); the National Center for Atmospheric Research Staff (Eds) for CFSR and CFSV2 are available at  (\url{https://rda.ucar.edu/#!lfd?nb=y&b=proj&v=NCEP\%20Climate\%20Forecast\%20System\%20Reanalysis}); the  SODA 3.3.1 products produced by Department of Atmospheric and Oceanic Science (\url{https://www.atmos.umd.edu/~ocean/}); the Brazilian National Water Agency from river flow data (\url{https://www.ana.gov.br/}); the Group for High Resolution SST (GHRSST) Regional/Global Task Sharing (R/GTS) framework from Operational SST and Sea Ice Analysis (OSTIA) system; the ETOPO1 Global Relief Model products produced by NOAA (\url{https://www.ngdc.noaa.gov/mgg/global/}); the altimeter products were produced by SSALTO/DUCAS and distributed by Copernicus Marine Environment Monitoring Service (\url{http://marine.copernicus.eu/services-portfolio/access-to-products/?option=com\_csw&view=details&product\_id=SEALEVEL\_GLO\_PHY\_L4\_REP\_OBSERVATIONS\_008\_047}); the OSU Tidal Data Inversion from TPXO Global Tidal Models (\url{https://www.tpxo.net/global}); the drifter data are available from the NOAA Global Drifter Program (\url{https://www.aoml.noaa.gov/phod/gdp/index.php}). 
The work of RD was performed in support of the US Department of Energy’s Fossil Energy, Oil and Natural Gas Research Program. It
was executed by NETL’s Research and Innovation Center, including work performed by Leidos Research Support Team staff under the RSS
contract 89243318CFE000003. 
This work was funded by the Department of Energy, National Energy Technology Laboratory, an agency of  the United States Government, through a support contract with Leidos Research Support Team (LRST). 
Neither the United States Government  nor any agency thereof, nor any of their employees, nor LRST, nor any of their employees, makes any warranty, expressed or implied, or assumes any legal liability or responsibility for the accuracy, completeness, or usefulness of any information, apparatus, product, or process  disclosed, or represents that its use would not infringe privately owned rights.
Reference herein to any specific commercial product, process, or service by trade name, trademark, manufacturer, or otherwise, does not necessarily constitute or imply its endorsement, recommendation, or favoring by the United States Government or any agency thereof. The views and opinions of authors expressed herein do not necessarily state or reflect those of the United States Government or any agency thereof.

%% file: sections/11_author-contributions.tex
M.B.G produced the results; M.B.G and R.D conducted the experiments; M.B.G, R.D and D.F.M.G wrote the paper; R.D ceded the code and assisted in the implementation;  R.T calculated EKE, MKE and TKE. All authors analysed the results and reviewed the manuscript.

%% file: sections/12_competing-interests.tex
The authors declare no competing interests.